# Broadband Dielectric Spectroscopy on dry Poly(vinyl alcohol)/Poly(vinylidene fluoride) blends reinforced with Nano - Graphene Platelets at combined pressure and temperature

Eirini Kolonelou, Anthony N. Papathanassiou[*]

National and Kapodistrian University of Athens, Physics Department. Panepistimioupolis, Section of Condensed Matter Physics, GR 15784, Zografos, Athens, Greece

[*] Corresponding author; email address: antpapa@phys.uoa.gr

**Abstract**

Poly(vinyl alcohol)/Poly(vinylidene fluoride) blends at mass ratio 3:1 with Nano - Graphene Platelet fillers constitute mechanically and thermally stable systems, which are used for developing piezoelectric devices. Blends host a fraction of water molecules absorbed by the poly(vinyl alcohol) phase. Dc conductivity and dielectric relaxation occur via fluctuation induced tunneling of electrons. Electric charge transport is affected by the glass transition of the polymer matrix and the rotational and translational dynamics of absorbed water molecules. In the present work, absorbed water was subtracted by annealing and pumping and, subsequently, dry blends were characterized by employing Broadband Dielectric Spectroscopy at combined temperature and pressure states. Conductivity and relaxation obey fluctuation induced tunneling temperature dependencies. The corresponding activation energies and activation volumes reveal the role of absorbed polar water molecules on the tunneling current for macroscopic or localized electron transport.

# 1. Introduction

Poly(vinylidene fluoride) (PVdF) is characterized by good mechanical properties, resistance to chemicals, high dielectric permittivity and pyro- and piezoelectric properties [1,2]. Piezoelectric PVdF is suitable for developing broadband acoustic and ultrasonic transducers [3] and medical imaging applications since they are flexible and with acoustic impedance similar to water and biological tissues [4]. The preparation of piezoelectric transducers based on PVdF requires the accurate characterization of the material properties [5, 6]. However, piezoelectric polymers exhibit intense electrical and mechanical losses compared to inorganic piezoelectric materials, emerging frequency, pressure and temperature studies of the complex dielectric permittivity $\varepsilon^*$. PVdF blended with poly(vinyl alcohol) (PVA) yield hybrid composite materials, which combine the physic – chemical characteristics of the end members. XRD, FTIR and Raman spectroscopy on PVA/PVdF of various mass fractions have proven that optimal structural and thermal stability and properties are attained at concentration 22.5 wt. % PVdF [7]

Graphene nano-fillers improve the electrical conductivity and thermal stability of PVA. Nano-graphene platelets (NGPs) dispersed in polymer matrix,v at mass fractions low enough to avoid a continuous percolation network through a network of physical touching NGPs. yield increased electrical conductivity of the composite. The percolation threshold for electric charge transport in polymer nano-composites loaded with various carbon allotropes are quite low [8, 9] and ranges from 0.1 to 0.3 % wt. % [8, 9, 13. Thermal fluctuations of the Fermi level of electron states in NGPs assist inter-NGP tunneling of electrons through the separating polymer barrier. Due to electrons tunneling, the effective percolation network consists of individual NGPs (or aggregates) and inter-NGP polymer volume through which tunneling current flows. As a result, he conductivity percolation cluster is formed at much lower fraction than that required for the formation of a continuous physical NGPs, due to the inter-NGP tunneling current through the polymer barrier.The tunneling current is tuned by the tunneling volume or the spacing between neighboring NGPs, the polarizability and thermodynamic state (glassy, semi-crystalline or rubber) of the polymer matrix, respectively. By monitoring how these parameters change upon compression, pressure sensing or pressure electro-switching polymer nano-composites can designed and produced. It is necessary to identify individual electric charge flow pathways and their relation with the polymer's glassy or rubber state, and kinetic phenomena of extrinsic entities, such as, absorbed water. Pressure-induced electro-switching has been seen in PVA/NGP composites as a result of competition between the components outlined above [9]. It is worth noticing that the piezoelectric nature of PVdF in PVdF/PVA compounds loaded with grapheme nanostructures have proved to be important for bio-medicine: The osteogenesis rate of human-induced pluripotent stem cells (hiPSCs) cultured on PVdF-PVA-graphene oxide (GO) electro spun scaffolds is significantly higher than other scaffolds, making them a highly promising scaffold-stem cell system for bone remodeling medicine [10].

PVA/PVdF (3:1 w/w) composites were prepared by drop casting a water solution of PVA and suspended PVdF micro-grains, reinforced with 2 wt. % NGPs. The blend composition is optimal to ensure good mechanical and thermal stability of the blend, which, has been used to develop hybrid piezoelectric devices [6]. The resulting free standing specimens are called as-received, or fresh. Differential Scanning Calorimetry (DSC), Scanning Electron Microscopy (SEM) and Broadband Dielectric Spectroscopy (BDS) were applied earlier at different temperatures and pressures [11]. Two processes affect the conductivity and relaxation mechanisms: the glass transition and a kinetic process due to the thermally activated mobility of water molecules absorbed by poly(vinyl alcohol. Electrical conductivity and relaxation studies at combined temperatures and pressures marked the shift of the glass transition temperature and the kinetic processes upon hydrostatic compression. In the present work, we performed Broadband Dielectric Spectroscopy on blends dried at 110$^o$ C in dynamic vacuum conditions. In Subsequently, Broadband Dielectric Spectroscopy was employed at combined temperature and pressure conditions. The application of hydrostatic pressure affects the structure and the electrical properties of the nano-composite in different ways, such as:

a)The separation between neighboring NGPs and, subsequently, tunneling length get shorter.
b) The polarizability of the polymer blend gets reduced and the inter -NGP tunneling current weakens.

c) The glass transition temperature changes monotonically with pressure; therefore, dc conductivity mechanisms are tuned by the modification of the polymer blend matrix from a rubber to a glassy state.
d) heterogeneity is likely to change due to small scale re-ordering of their elements, so as to make better use of the volume of the compressed material.
(v) The volume fraction of the conductive inclusions increases on hydrostatic compression as a result of the reduction of the specimen's volume. Subsequently, the percolation threshold determined at ambient conditions, is supressed, too.

## 2. Materials and methods

Free standing specimens were prepared by drop casting a water solution of polyvinyl alcohol (PVA) (purity 99,5 %; ASG Scientific, CAS 9002-89-5) with dispersed micro-granules of poly(vinylidene fluoride (PVdF) of molecular weight 534.000 (Sigma Aldrich; CAS 24937-79-9) graphene nano-platelets (NGPs) (Angstrom Materials Ltd. The PVdF mean grain size 2.5 μm [12]. According to the manufacturer, the lateral size of NGPs is ≤ 10 μm$^2$ and the average through dimensions were 50 – 100 nm determined by BET surface analysis and size distribution. The NGPs mass fraction was 2 wt. %, slightly exceeds the percolation threshold [13], to prevent the formation of a continuous network of NGPs. The homogeneous distribution of NGPs and the absence of any continuous NGPs network of NGPs were verified by SEM imaging [7]. SEM, DSC, electromechanical coupling characterization and BDS at ambient pressure and various temperatures have been reported in refs [7, **11**]. The mass fraction of NGPs was 2 wt %, which is certainly above the critical charge percolation threshold (0.1 – 0.3 wt %) and is practically unaffected from volume changes induced by pressure or temperature.

PVA accommodates water molecules as a result of drop casting of its water solution suspending PVDF micro-granules and NGPs at ambient conditions of temperature, pressure and humidity. In the present work, blends were dried at 110$^o$ C inside a chamber connected to a rotary pump. The time duration of drying ranged from 1 to 24 hours and concluded that 1 hour treatment was adequate to extract the maximum quantity of absorbed water. The water content, estimated from the relative change in the mass of the specimen after drying, was ~ 2.3 % wt of the total mass of the as-received blend.

A Novocontrol High Pressure apparatus, suitable for Broadband Dielectric Spectroscopy was used to study the frequency dependence of the dielectric properties of the blends at various temperatures and pressures. The tangent of the loss factor tanδ and complex electric modulus M* were measured in the frequency range from 10 mHz to 1 MHz, by employing a Solartron 1204 Impedance Analyser combined with a Broadband Dielectric Converter (BDC, Novocontrol). Measurements were carried out at combined isothermal and isobaric conditions. Temperature increased from T=20$^o$C to 120$^o$C at a step of 10 degrees. Pressures increased by 250 bars from 1 bar to 1750 bar.

## 3. Theoretical background

The dc conductivity of polymers and polymer blends is improved significantly upon filling with NGPs [8]. The electronic density is determined by the free electrons available by NPGs. Dc conductivity increases systematically upon mass fraction of NGPs and reaches its percolation threshold around 0.3 wt % NGPs [8]. A mass fraction of 2 wt % ensures that the dispersed NGPs provide high density of electron states. The activation energy values are of the order of a few meV, i.e., orders of magnitude smaller than the band gap of an insulator (polymer). Electrons, rather than ions, can tunnel through the potential barrier dictated by the polymer separating neighboring NGPs by expensing energy of a few meV). Possibly, ionic transport of external defects or protonic conductivity are likely to occur, but are weak compared with the electron tunneling process.

Sheng proposed the Fluctuation Induced Tunneling (FIT) theory to explain electronic transport in heterogeneous materials consisting of an insulating matrix and dispersed conducting inclusions [14]. FIT explained initially electronic transport in granular metals as inter-grain electron tunneling assisted by thermal fluctuations of their Fermi energy level. FIT described effectively to a wide range of electron-conducting in homogeneously disordered materials, including conducting polymers and composites. The temperature dependency of dc conductivity:

$$\sigma_{dc} = \sigma_0 exp\left(\frac{-T_1}{T+T_0}\right) \qquad (1)$$

According to Sheng and Klafter [15, 17] a $T^{-1/2}$ dependence applies, as well:

$$\sigma_{dc} = \sigma'_0 exp\left(-\frac{\mathcal{T}_1}{T}\right)^{1/2} \qquad (2)$$

$T_1$, $T_0$ and $\mathcal{T}_1$ denote the fitting parameters of the temperature dependence of the dc conductivity.

Atomic charge flow events underlie both (macroscopic) dc conductivity and spatially localized charge transport. If $\Gamma_{ij}$ denotes the transition rate of electric charges tunneling from site i to site j, in the presence of an external field, then, $\sigma_{dc} \propto \Gamma_{ij}$ and $\tau^{-1} \propto \Gamma_{ij}$, where $\tau$ is the dielectric relaxation time. In the frequency domain, dielectric loss maximizes at $f_{max} = 1/\tau$, thus: $f_{max} \propto \Gamma_{ij}$. As a result, the temperature dependence of the relaxation maximum resembles that of $\sigma_{dc}$, i.e.:

$$f_{max} \propto exp\left(\frac{-t_1}{T+t_0}\right) \qquad (3)$$

and approximately:

$$f_{max} \propto exp\left(-\frac{t_1}{T}\right)^{1/2} \qquad (4)$$

$t_1$, $t_0$ and $\mathbf{t}_1$ denote the fitting parameters of the temperature dependence of the relaxation frequency. The order of magnitude of the effective potential barrier is $k_B \mathcal{T}_1$ for dc conductivity and $k_B \mathbf{t}_1$ for dielectric relaxation. A $T^{-1/2}$ dependence resemble that predicted by Mott's Variable Range Hopping (VRH) model, but has different physical basis: FIT accounts for inhomogeneous disordered materials whereas electron states extend over the volume of individual conducting grains embedded in an insulating matrix and tunneling is induced by thermal fluctuations of the Fermi level. VRH accounts for phonon assisted tunneling of electrons from a localized state to another one within a homogeneously disordered solid. The use of eqs. (2) and (4) is more advantageous (compared to eqs. (1) and (3), respectively) to fit the experimental data points contain a single parameter to be determined.

## 4. Results and Discussion
### 4.1 Study of the dc conductivity by means of complex electric modulus

The complex electric modulus $M^* = M' - iM''$ is defined as: $M^* \equiv \frac{1}{\varepsilon^*}$, where:

$$M'(f) = M_s \frac{(f\tau_\sigma)^2}{1+(f\tau_\sigma)^2} \qquad (5)$$

$$M''(f) = M_s \frac{f\tau_\sigma}{1+(f\tau_\sigma)^2} \qquad (6)$$

where $i^2 \equiv -1$ and $\tau_\sigma$ is a characteristic relaxation time. Macroscopic dc conductivity contributes a low frequency "conductivity peak" in the $M''(f)$ spectra, which has its maximum at a frequency $f_{max,\sigma} = (2\pi\tau_\sigma)^{-1}$, or:

$$f_{max,\sigma} = \sigma_{dc}/(2\pi\varepsilon_0\varepsilon_\infty) \qquad (7)$$

where $\varepsilon_0$ denotes the permittivity of free space and $\varepsilon_\infty \equiv \varepsilon'(f \to \infty)$. $\varepsilon'(f \to \infty)$ has usually a weak temperature dependence, which is confirmed in our experiments, too. As $f_{max,\sigma}$ and $\sigma_{dc}$ practically share a common temperature dependence and, hence, Eqs. (3) and (4) are suitable to fit $f_{max,\sigma}(T)$ data sets. At higher frequencies than those detected in $tan\delta(f)$, dielectric relaxation peaks also appear in the $M''(f)$ spectra [16, 17]. The electric modulus function, defined as the inverse of complex permittivity, supresses the undesirable low frequency electrode polarization capacitance, allowing a precise estimate of the activation energy of dc conductivity [18, 19].

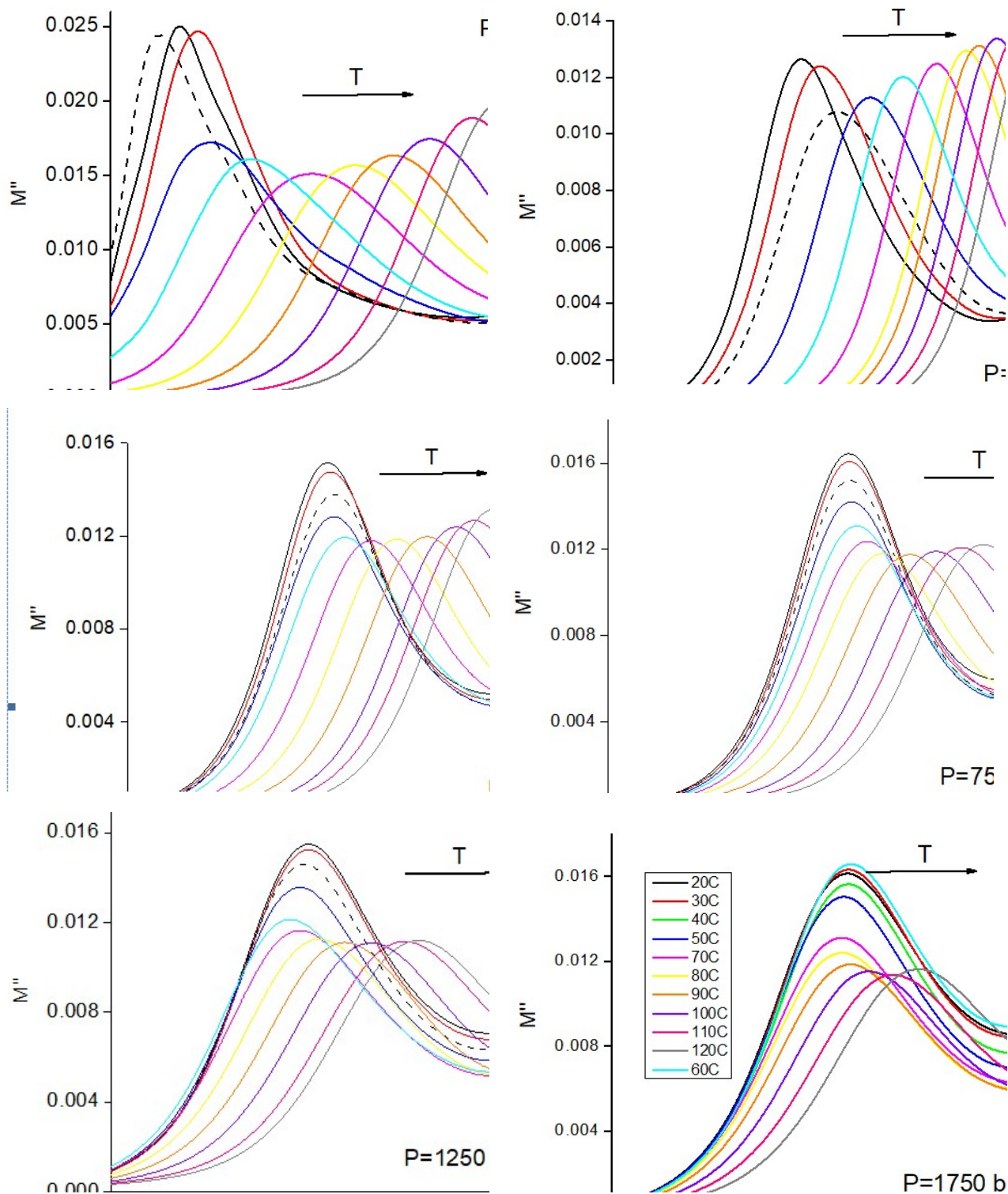

***Fig. 1.*** *Typical isotherms of the imaginary part of the electric modulus M′′ as a fu of the working frequency f, recorded at different pressures. Temperature ranged from 20° C to 120°C at intervals of 10 degrees.*

Typical plots of the imaginary part of the complex electric modulus $M''(f)$ for various temperature and pressure valuaes are depicted in Fig. 1. The temperature dependence of the maximum Comparative representation of the temperature dependences of the logarithm of the maximum $f_{max,\sigma}$(of he conductivity peaks for a couple of pressure values, are presented in Fig. 2. We observe that the high temperature data points of $logf_{max,\sigma}(T^{-1/2})$ can be fitted by straight lines, according to Eq. (2). Below a critical temperature (marked up by an arrow), below which, the polymer matrix is in its glassy state; i.e., the arrow signs the glass transition temperature as a change in the monotony of the $logf_{max,tan\delta}(T^{-1/2})$ diagram. We note that this characteristic temperature traced in dielectric spectra at P=1 bar, is compatible with the glass transition temperature determined (at ambient pressure) by Differential Scanning Calorimetry (DSC) [7]. As can be seen in Fig. 2, the transition temperature shifts towards higher temperatures upon increasing pressure. The pressure derivative of the glass transition temperature signed through dielectric spectroscopy is estimated around $dT_g/dP$=0.030 grad/bar, which is compatible with a mean typical value of 0.04 grad/bar for polymers [20]. The fitting parameter $T_1$ of Eq. (4) yields an estimate of the activation energy for dc conductivity $E_{M''} \approx k_B T_1$. The latter is interpreted as an effective potential barrier, which is depicted as a function of pressure in Fig. 3. The activation energy values $E_{M''}$ distribute from $E_{M''}(P = 1bar) = 51(2)meV$ to $E_{M''}(P = 1750bar) = 24(3)meV$. For as-received samples (containing water molecules absorbed by the PVA phase), a couple of conductivity mechanisms operating at low pressures, which merge to a single one above 1000 bar $E_{M''}(P = 1bar) = 78(5)meV$ to $E_{M''}(P = 1750bar) = 80(3)meV$. The removal of water molecules affects the modes for dc conductivity and reduces the energy cost required for tunneling. Macroscopic charge transport occurs by successive inter-grain fluctuation induced tunneling through the polymer separating neighboring NGPs. However, the internal electric field and the electric force along the direction of the external electric field, decreases (compared to the externally applied field) upon the polymer's polarizability. Since the polarizability of neat PVA is smaller than that of the hydrated one, the activation energy values (which measure of the energy cost required for a single electron fluctuation induced tunneling) in dry blends are lower than those of as-received ones, respectively.

The activation volume for a thermally activated process is thermodynamically defined as: $v^{act} \equiv \left(\frac{\partial g^{act}}{\partial P}\right)_T$, where $g^{act}$ denotes the Gibbs free energy per transition [9, 21]. To a first approximation, if the entropy change per transition has weak dependence upon pressure, $\left(\frac{\partial g^{act}}{\partial P}\right)_T \cong \left(\frac{\partial E_{M''}}{\partial P}\right)_T$. Thus:

$$v^{act}_{M''} \cong \left(\frac{\partial E_{M''}}{\partial P}\right)_T \qquad (8)$$

$v^{act}_{M''}$ is a measure of the effective potential barrier ($\sim k_B T_1$) modification upon pressure and the fluctuation of the macroscopic volume per individual flow event. The potential barrier decreases on increasing from 1 to 750 bar: the slope of a straight line fitted to the data points of this region provides (according to Eq. (8)) the activation volume for dc conductivity through the pressure evolution of the 'conductivity peak': $v^{act}_{M''} = -62(6)\text{Å}^3$. The negative sign indicates that pressure, by supressing the specimen's volume, reduces the tunneling range. Beyond 750 bars, $E_{M''}$ reached a near-saturation value. The fact that $E_{M''}$ is low enough (i.e., a few tens of meV) and the negative sign in the corresponding effective activation volume confirm that the dominant macroscopic conductivity involves electron tunneling. The linearity observed in the $logf_{max,\sigma}(T^{-1/2})$ plots evidence for a single FIT conductivity mechanism. A change in the monotony of $E_{M''}(P)$ is therefore visualized as a change in the magnitude of the volume fluctuations as the polymer achieves a highly packed state in excess of its initial free volume one.

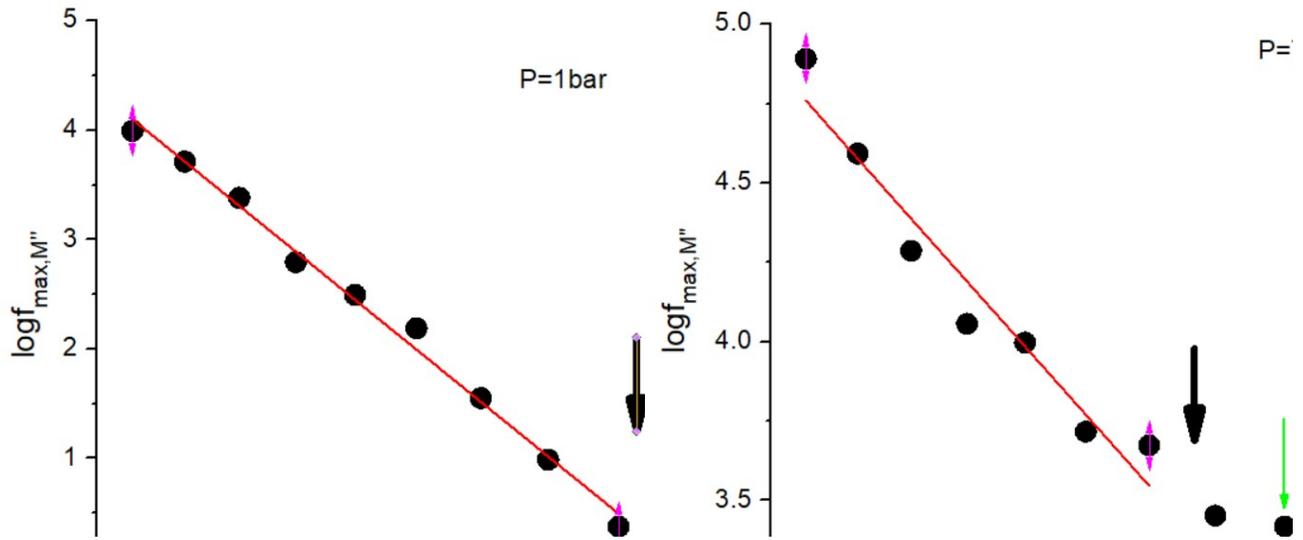

***Fig. 2.*** *Temperature dependence of the logarithm of the maximum of the M´´(f) conductivity peaks $logf_{max,\sigma}(T^{-1/2})$, at two different pressure values. Straight lines are best fits of eq. (2) to the higher temperature data. Arrows mark the glass transition temperature whereas the monotony of $logf_{max,tan\delta}(T^{-1/2})$ changes. The pressure derivative of the glass transition temperature signed through dielectric spectroscopy is estimated around $dT_g/dP=0.030$ grad. Arrows mark the glass transition temperature at different pressures.*

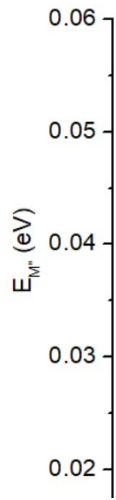

***Fig. 3.*** *The activation energy of the 'conductivity peak' appearing in $M''(f)$ spectra $E_{M''} \approx k_B T_1$, where $T_1$ is the fitting parameter of Eq. (4). A straight line fits the low pressure data points.*

**4.2 Study of relaxation of trapped electric charge by means of the loss angle tangent**

Isotherms of the tangent of the loss angle as a function of frequency collected from T=20°C to 120°C by a step of 10 grads at ambient pressure for as are depicted in Fig. 4. Dielectric relaxation peaks, overlapping with a dc conductivity baseline, get prominent on increasing temperature (Fig. 4). Dielectric relaxation appears upon filling of the polymer blend with NGPs [7]. In Fig. 5a, a typical isobaric *logf$_{max,tanδ}$(T $^{-1/2}$)* diagrams is presented. Data points are fitted by a single straight line, according to eq. (4), from which, the fitting parameter $t_1$ yields the activation energy $E_{tanδ} \approx k_B t_1$. The activation energy values evaluated at different pressures are presented in Fig. 5b. The $E_{tanδ}$ values are of the order of a few tens of meV, which - in combination with the fact that the fluctuation induced tunneling (FIT) describes *logf$_{max,tanδ}$(T $^{-1/2}$)* successfully (Fig. 5b), - ensures that relaxation involves electron tunneling. We note that the activation energy values are comparable to those reported for the as-received blend hosting water molecules (such as $E_{tanδ}(P = 1bar) = 60(2)meV$) [**11**]. The latter conclusion is additionally supported by the negative sign of the activation volume of relaxation $v^{act}_{tanδ}$, for which, a relation analogous to Eq. (8) can be written:

$$v^{act}_{tanδ} \cong \left(\frac{\partial E_{tanδ}}{\partial P}\right)_T \qquad (9)$$

The slope of a straight line fitted to the $E_{tanδ}(P)$ data points of Fig. 5b, yields, through Eq. (9): $v^{act}_{tanδ} = -24(6)Å^3$, which, in turn, implies that pressure favors the extension of electron wave functions and assists electron relaxation through tunneling. Relaxation may occur within NGPs (or NGP aggregates) or by localized inter-NGP fluctuation induced tunneling through the polymer spacing. As NGPs are harder than the polymer matrix of the blend, the distance between neighboring NGPs (and, hence, the inter-NGP tunneling range) is strongly affected by hydrostatic compression, compared with the weak volume change of each NGP. Subsequently, relaxation is probably attributed to inter-NGP relaxation of electrons.

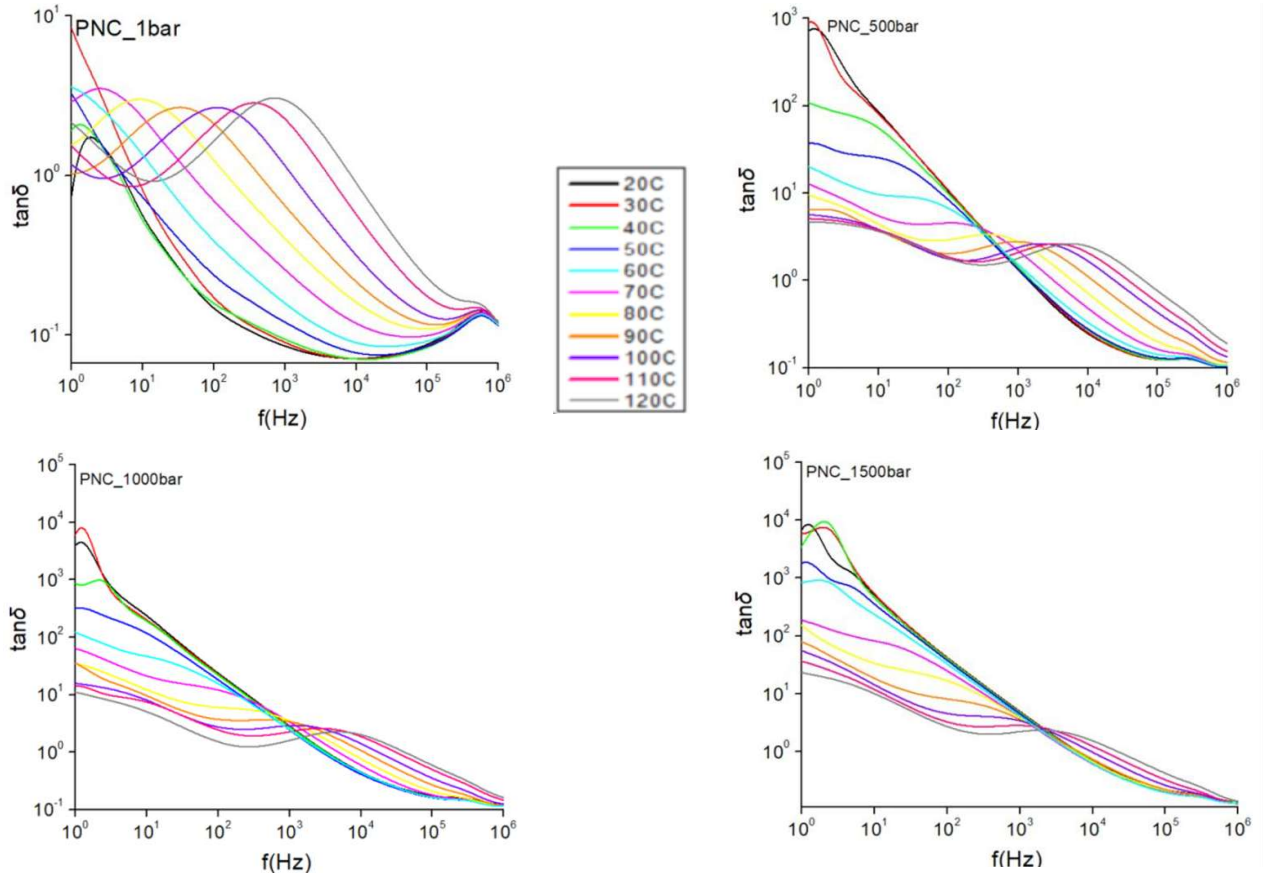

*Fig. 4. Typical isotherms of the tangent of the loss angle tanδ as a function of frequency f, collected at various pressures, for dry specimens.*

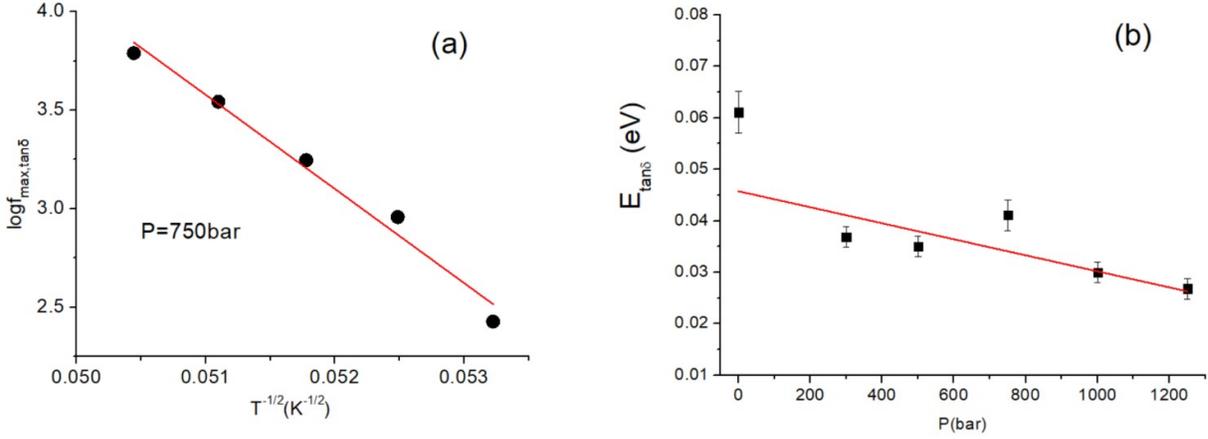

**Fig. 5.** *(a) A typical plot of $\log f_{max,\tan\delta}(T^{-1/2})$ for the dielectric relaxation mechanism observed in the $\tan\delta(f)$ spectra, at 750 bar, above the glass transition temperature. A straight line fits the data points, according to eq. (3). (b) The activation energy $E_{\tan\delta} \approx k_B t_1$, where $t_1$ is the fitting parameter of Eq. (4). The straight line fits the data points.*

In polar polymers with nano-carbon fillers, the effect of hydrostatic pressure on the inter-NGP penetrating quantum mechanically the potential battier set by the polymer separation, is tuned by two competing phenomena:
(i) reduction of the tunneling length, which enhances the tunneling current,
(ii) reduction of the polarizability of the polymer (through which, electrons tunnel), which impedes tunneling [22 9, 21, 11].

The polar molecules of water absorbed in as-received specimens, magnify the overall polarizability of PVA system: it was found that the pressure dependence of the polarizability dominated that of tunneling length shortening, yielding an increase of the effective potential barrier or the effective activation energy $E_{\tan\delta}(P)$ [**11**]. As a result, according to Eq. (9), $\upsilon_{\tan\delta}^{act}$ was found positive: [**11**] $\upsilon_{\tan\delta}^{act} = +26(6)\text{Å}^3$. By comparing with the value $-24(6)\text{Å}^3$ estimated in the present work, we observe that the both activation volumes share practically the same absolute value, indicating that relaxation in as-received and dry blends are of the same nature and induce the same volume fluctuation. The positive sign of the activation volume in as-received blends probably stems from the fact that the overall polarizability of the hydrated polymer blend is larger than that of the dry one: the competition between phenomenon (i) and (ii), define the pressure evolution of $E_{\tan\delta}(P)$ and the sign of $\upsilon_{\tan\delta}^{act}$.

## 5. Conclusions

Poly(vinyl alcohol)/Poly(vinylidene fluoride) blends at mass ratio 3:1 with Nano - Graphene Platelet fillers were characterized by employing Broadband Dielectric Spectroscopy. Electric charge transport is affected by the glass transition of the polymer matrix and the rotational and translational dynamics of water molecules absorbed by poly(vinyl alcohol). In the present work, absorbed water was subtracted by annealing and pumping and, subsequently, dry blends were characterized by applying Broadband Dielectric Spectroscopy at combined temperature and pressure states.

A single dc conductivity operates above the glass transition temperature. The activation energy determined at different pressures indicate are lower than those reported for the as-received blend. Dry specimens exhibit lower polarizability than hydrated blends, hence, the internal electric field (in turn, the magnitude of electric force applied along the tunneling current) is larger than that in hydrated blend of higher polarizability. Thus, the energy required for electron tunneling is lower in dry blends.The activation energy is a decreasing function of pressure, as a result of shortening of the inter-NGP tunneling range upon hydrostatic compression.

A dielectric relaxation mechanism appearing in the tangent loss angle representation is described by the fluctuation induced tunneling model and involves inter-NGP electron trapping. The relaxation activation volume shares practically the same absolute value with that reposted for as-received blends, indicating that relaxation in as-received and dry blends are of the same nature and induce the same volume fluctuation. The competition between the reduction of the tunneling length, which enhances the tunneling current, and the reduction of the polarizability of the polymer (through which , electrons tunnel), which impedes tunneling, define the pressure evolution of the activation energy and the sign of the corresponding activation volume.

# References

[1] M.G. Broadhurst, G.T. Davis, J.E. McKinney, R.E. Collins, Piezoelectricity and pyroelectricity in polyvinylidene fluoride - A model, J. Appl. Phys. 49 (1978) 4992.

https://doi.org/10.1063/1.332119.

[2] D.K. Das-Gupta, On the nature of pyroelectricity in polyvinylidene fluoride, Ferroelectrics 33 (1981) 75.

https://doi.org/10.1080/00150198108008072.

[3] G. R. Harris, R. C. Preston, and A. S. DeReggi, The impact of piezoelectric PVDF on medical

ultrasound exposure measurements, standards, and regulations, IEEE Trans. Ultrason., Ferroelect., Freq. Contr. 47 (2000): 1321.

https://doi.org/10.1109/58.883521

[4] F. Stuart Foster, K. A. Harasiewicz, M. D. Sherar, A History of Medical and Biological Imagingwith Polyvinylidene Fluoride (PVDF) Transducers, IEEE Trans. Ultrason., Ferroelect., Freq. Contr. 47 (2000): 1363.

https://ieeexplore.ieee.org/document/883525.

[5] W. P. Mason, Technical Digests: An Electromechanical Representation of a Piezoelectric Crystal Used as A Transducer, Bell System Technical Journal 14 (1935) 718.

https://doi.org/10.1002/j.1538-7305.1935.tb00713.x.

[6] D. Valadorou, A.N. Papathanassiou, E. Kolonelou, E. Sakellis, Boosting the electro-mechanical coupling of piezoelectric polyvinyl alcohol–polyvinylidene fluoride blends by dispersing nano-graphene platelets, J. Phys. D: Appl. Phys. 55 (2022) 295501.

https://doi.org/10.1088/1361-6463/ac629d.

[7] E. Kolonelou , E. Loupou , P. A. Klonos , Elias Sakellis , D. Valadorou , A. Kyritsis b, A. N. Papathanassiou, **Thermal and electrical characterization of poly(vinyl)alcohol)/poly (vinylidene fluoride) blends reinforced with nano-graphene platelets**, Polymer (2021) 123771.

https://doi.org/10.1016/j.polymer.2021.123731

[8] Ch. Lampadaris, I. Sakellis and A. N. Papathanassiou, **Dynamics of electric charge transport and determination of the percolation insulator-to-metal transition in polyvinyl-pyrrolidone/nano-graphene platelet composites,** Appl. Phys. Lett. 110 (2017) 222901.

 https://doi.org/10.1063/1.4984203.

[9] E. Kolonelou, A.N. Papathanassiou, E. Sakellis, **Evidence of local softening in glassy poly(vinyl alcohol)/poly(vinyl pyrrolidone) (1/1, w/w) nano-graphene platelets composites,** Materials Chemistry and Physics 223 (2018) 140-144.

 https://doi.org/10.1016/j.matchemphys.2018.10.058.

[10] E. Azadian, et al., Polyvinyl alcohol modified polyvinylidene fluoride-graphene oxide scaffold promotes osteogenic differentiation potential of human induced pluripotent stem cells, Journal of Cellular Biochemistry 121 (2020) 3185-3196.

https://doi.org/10.1002/jcb.29585.

[11]    Kolonelou E., Loupou E., Sakellis E., Papathanassiou A.N., Pressure and temperature dependence of the electric modulus and loss factor of Poly(vinyl alcohol)/Poly(vinylidene fluoride) blends reinforced with Nano - Graphene platelets, J. Phys. Chem. Solids,2023 178,2023, 111277

https://doi.org/10.1016/j.jpcs.2023.111277.

[12] H. Kim, et al, Fabrication of bulk piezoelectric and dielectric BaTiO3 ceramics using paste extrusion 3D printing technique. Journal of the American Ceramic Society.

102 (2018).

 https://doi.org/10.1111/jace.16242.

[13] G. Krishna E. Bama, E.K. Indra Devi, **Structural and thermal properties of PVdF/PVA blends** J Mater Sci 44 (2009)  1302–1307.

https://doi.org/10.1007/s10853-009-3271-8.

[14] P. Sheng, J. Klafter, Hopping conductivity in granular disordered systems, Phys. Rev. B 27 (1983) 2583–2586.

https://doi.org/10.1103/Phys

[15] P. Pissis, A. Kyritsis, Hydration studies in polymer hydrogels, J. Polym. Sci. B Polym. Phys. 51 (2013) 159–175.

https://doi.org/10.1002/polb.23220.

[16] A. Wurm, M. Ismail, B. Kretzschmar, D. Pospiech, C. Schick, Retarded crystallization in polyamide/layered silicates nano-composites caused by an immobilized interphase, Macromolecules 43 (2010) 1480–1487.

https://doi.org/10.1021/ma902175r.

[17] R. Bosisio, C. Gorini, G. Fleury, J.-L. Pichard, Gate-modulated thermopower of disordered nanowires: II. Variable-range hopping regime, New J. Phys. 16 (2014) 095005.

https://doi.org/ 10.1088/1367-2630/16/9/095005.

[18] A.N. Papathanassiou, O. Mykhailiv, L. Echegoyen, I. Sakellis, M. Plonska- Brzezinska, Electric properties of carbon nano-onion/polyaniline composites: a combined electric modulus and ac conductivity study, J. Phys. D Appl. Phys. 48 (2016) 285305.

https://doi.org/ 10.1088/0022-3727/49/28/285305

[19] P.S. Das, P.K. Chakraborty, B. Behera, R.N.P. Choudhary, **Electrical properties of $Li_2BiV_5O_{15}$ ceramics**, Phys. B: Condens. Matter, 395 (2007), 98-103
https://doi:10.1016/j.physb.2007.02.065

[20] Belfiore, Physical Properties of Macromolecules, Wiley Publications (2010) p. 177.

[21] Eirini Kolonelou, Anthony N. Papathanassiou, Sakellis Elias, Pressure-induced Electro-Switching of Polymer/nano-Graphene Composites Materials Chemistry and Physics, 232 (2019) 319–324, https://doi.org/10.1016/j. matchemphys.2019.05.002.

[22] Syurik, J.; Ageev, O.; Cherednichenko, D.; Konoplev, B.; Alexeev, A. Non-linear conductivity

dependence on temperature in graphene-based polymer nanocomposite, Carbon, 2013, 63, 317-323.

http://dx.doi.org/10.1016/j.carbon.2013.06.084.